\documentstyle[preprint,eqsecnum,aps,epsfig]{revtex}

\tightenlines

\input epsf.sty

\newcommand{\be}{\begin{equation}}    
\newcommand{\ee}{\end{equation}}
\newcommand{\beq}{\begin{eqnarray}}
\newcommand{\eeq}{\end{eqnarray}}
\newcommand{\beqn}{\begin{eqnarray*}}
\newcommand{\eeqn}{\end{eqnarray*}}
\newcommand{\f}[2]{\frac{#1}{#2}}

\newcommand{\dps}{\displaystyle}        

\def\op{ \ $ }
\def\cl{$ \ }
\def\nn{\nonumber}

\def\gappreq{\!\stackrel{\scriptscriptstyle >}{\scriptscriptstyle \sim}\!}
\def\ver{\vskip 12pt}
\def\ii{{\rm i}}   

\def\N{N_{l m}}
\def\T{T_{l m}}
\def\V{V_{l m}}
\def\L{L_{l m}}

\def\hh{h^{0}_{l m}}
\def\hs{h^{1}_{l m}}
\def\pps{\Psi_{l m}}                                                                      
\def\IL{\relax{\rm I\kern-.18em L}}

\def\o{\omega}

\begin{document}


\draft

\title{Gravitational signals emitted by a  point mass orbiting a neutron
star: effects of stellar structure.}

\author
{J.A. Pons$^{1}$, E. Berti$^{2}$, L. Gualtieri$^1$, G. Miniutti$^{1}$
and V. Ferrari$^{1}$}
\address
{$^1$ Dipartimento di Fisica ``G.Marconi",
 Universit\` a di Roma ``La Sapienza"\\
and Sezione INFN  ROMA1, piazzale Aldo  Moro
2, I-00185 Roma, Italy\\
$^2$ Department of Physics, Aristotle University of Thessaloniki,
Thessaloniki 54006, Greece}

\date{\today}

\maketitle

\begin{abstract}
The effects that the  structure of a neutron star  would have  
on the gravitational emission of a binary system 
are studied in a perturbative regime, and in the frequency domain.
Assuming that a neutron star is perturbed by a point mass  moving
on a close, circular orbit, we solve the equations of stellar 
perturbations in general relativity to  evaluate
the energy lost by the system in gravitational  waves.
We compare the  energy output obtained for different stellar models
with that found by assuming that the perturbed object is a black hole 
with the same mass, and we discuss the role played by the excitation of the stellar
modes. 
Our results indicate that the stellar structure  begins to affect
the emitted power  when the orbital velocity is 
$v\gappreq 0.2~c$ ($\nu_{GW}\gappreq 185$ Hz for a binary system composed of two
$1.4 M_\odot$ neutron stars). We show that the  differences between different
stellar models and a black hole are due mainly  to the 
excitation of the quasinormal modes of the star.
Finally, we discuss to what extent and up to which distance the perturbative
approach can be used to describe the interaction of a star and a 
pointlike massive body.

\end{abstract}
 

\narrowtext

\section{Introduction}

\ver\ver
In a recent paper\cite{tutti1} (to be referred to hereafter as Paper I)
we have studied the gravitational emission of a binary system by using the following
perturbative approach:   one of the two stars  is assumed to be
an extended body, whose equilibrium structure is described by
an exact solution of the relativistic equations of  hydrostatic equilibrium;
the second star is a pointlike mass which  induces
a perturbation on the gravitational field and on the
thermodynamical structure of the extended companion. 
We modeled the extended star using  a polytropic equation of state,  we
chose the parameters so that  the radius and the mass  were those of a plausible
neutron star (NS), and we solved
the equations of stellar perturbations to compute
the power emitted in gravitational waves when the point
mass moves  on  orbits of arbitrary eccentricity.

In this paper we use the same perturbative approach to compute the power radiated 
in gravitational waves by  different stellar models.
The  purpose of this investigation  is to answer a number of questions that arise 
in studying the signals emitted by NS-NS binary systems
during the last phases before coalescence.
The first is about the role played by the internal structure of the 
star: how does the gravitational emission depend on the mass and radius of the
star (and consequently on the equation of state (EOS) of dense matter)? 
In particular,  what is the difference if the perturbed object
is a black hole (BH)?
The second is related to the possibility of exciting the quasi normal 
modes of the star. Are the differences between stars and a black hole
(if any) due exclusively to resonant modes or to other orbital effects?
The third question concerns  the  domain of applicability of
the perturbative approach: to what extent are the results of this approach
applicable to describe the evolution of a true binary system, composed of two 
neutron stars of comparable mass?
Are we still in a perturbative regime when the two neutron stars are, say, 
3 or 4 stellar radii apart?

The plan of the paper is the following.
In Sec. II  we shortly review  the procedure we use to find the solution of the 
equations of stellar perturbations inside and outside the perturbed star, 
along the lines of Paper I. Since we are mainly interested in the last 
phases of the orbital evolution, we
shall consider the orbit of the point mass as already circularized \cite{peters}.
In Sec. III  we discuss in detail the results of the integration of the perturbed
equations for the different stellar models and for a black hole, both excited by 
the same process.
In Sec. IV we examine the domain of applicability  of the perturbative approach
applied to binary systems, and in Sec. V we draw the conclusions.

\section{The perturbed equations}

In order to compute the radiation emitted by a  a neutron star perturbed
by a massive  point particle, we need to integrate the equations of stellar
perturbations inside and outside the star, and  this can be accomplished by using
different gauges and formalisms. Inside the star we choose  the
Chandrasekhar-Ferrari gauge \cite{chandrafer1}, which allows to decouple the
equations for the gravitational perturbations from those describing the 
perturbations of the fluid.  After expanding the perturbed metric tensor
in tensorial spherical harmonics,
choosing this gauge  the radial part of the {\it polar} metric perturbations
is described by four functions,
$\Bigl[\N(\omega,r), \L(\omega,r), \V(\omega,r),\T(\omega,r)\Bigr],$
whereas only two functions, $\Bigl[  \hh(\omega,r),\hs(\omega,r)\Bigr],$ are needed to
describe the {\it axial} part.
As in Paper I,  we integrate the equations of stellar perturbations in the frequency domain
(cf. Paper I, Eqs. (2.2) and (2.4)) up to the surface of the star, where
we construct the Zerilli and the Regge-Wheeler
functions, \op Z^{pol}_{l m}(\omega,r)\cl and \op Z^{ax}_{l m}(\omega, r)\cl    
\cite{zerilli},\cite{reggewheeler}
\beq
\label{zerfun}
Z^{pol}_{l m}(\omega, r)&=&\f{r}{nr+3M}\Bigl[
3M V_{l m}(\omega,r)-rL_{l m}(\omega,r)
\Bigr],\\\nn
Z^{ax}_{l m}(\omega, r)&=&- \frac{e^{2\nu}}{r}
h^1_{l m}(\omega, r),
\eeq
and their first derivatives with respect to $r$. 
The integration of the equations of stellar perturbations in the interior
(together with the TOV equation) is done using an adaptive step Runge-Kutta 
method. The relative tolerance of the integrator is set to $10^{-8}$.
We use as integration variable the logarithm of the pressure, which allows
us to approach the stellar surface gradually, and to determine
the radius with the required accuracy. 

The two functions in (\ref{zerfun}) are needed
to compute  the radial part, $\pps  (\omega, r)$, 
of the perturbation of the Weyl scalar \op \delta\Psi_4,\cl  
which is defined as
\be
\label{psiquattro}
\pps (\omega, r)=\frac{1}{2\pi}
\int
d\Omega~dt~e^{ i\omega t}~_{-2}S^\ast_{l m}(\theta,\phi)
\left[ r^4~\delta\Psi_4(t,r,\theta,\phi)\right],
\ee
where \op _{-2}S_{l m}(\theta,\phi)\cl is the
spin-weighted spherical harmonic of spin $-2$.
In terms of 
\op Z^{pol}_{l m}(\omega,r)\cl and \op Z^{ax}_{l m}(\omega, r),\cl 
\op \pps (\omega,r)\cl is
\beq
\label{relazione1}
\pps(\omega,r)&=&\frac{r^3\sqrt{n\left(n+1\right)}}{4\omega}\left[
V^{ax} Z^{ax}_{l m}  +\left(W^{ax} +2i\omega\right)\Lambda_+ Z^{ax}_{l m}  \right]\\
\nn
&-&\frac{r^3\sqrt{n\left(n+1\right)}}{4}\left[
V^{pol} Z^{pol}_{l m}+\left(W^{pol}+2i\omega\right)\Lambda_+  Z^{pol}_{l m}
\right], \eeq
where \op 2n=(l-1)(l+2),\cl
\op
\Lambda_+={d\over dr_*}+i\omega=
{\Delta\over r^2}{d\over dr}+i\omega,
\cl
\op V^{pol}\cl and \op V^{ax}\cl are the  Zerilli and the Regge-Wheeler potentials
\cite{zerilli},\cite{reggewheeler}, 
and
\beqn
W^{ax}&=&{2\over r^2}(r-3M),\\
W^{pol}&=&2{nr^2-3Mnr-3M^2\over r^2(nr+3M)}.
\eeqn
Outside the star we integrate the inhomogeneous  Bardeen-Press-Teukolsky
BPT equation \cite{bardeenpress,teukolski}
\be
\label{teukolsky}
\left\{\Delta^2\frac{d}{dr}\left[\frac{1}{\Delta}\frac{d}{dr}\right]+
\left[\frac{\left(r^4\omega^2+4i(r-M)r^2\o\right)}{\Delta}
-8i\o r-2n\right]\right\}\pps (\o, r)= -T_{l m}(\o, r),
\ee
where \op \Delta=r^2-2Mr,\cl
and the source term \op T_{l m}(\o, r)\cl is that appropriate to describe the point
mass  \op m_0\cl moving on a given orbit around the star. In Paper I we discussed
how to construct the solution of eq. (\ref{teukolsky}) in the general case of
elliptic orbits. In this paper, since we are interested mainly in the last phases
of the evolution of binary systems, when the orbit has already been circularized,
we focus on circular orbits and give the explicit solution of the equations in that
case.
If the mass \op m_0\cl moves on an orbit of radius $R_0$
the geodesic equations  give
\beq
\bar{\gamma}\equiv\dot t = \frac{E}{1-\frac{2M}{R_0}}, \qquad
\omega_K\equiv{d\varphi\over dt}={\dot{\varphi}\over\bar{\gamma}}\,,
\eeq
where the dot indicates differentiation with respect to proper time.
$E$ is the energy of the particle per unit mass, and
$\omega_K$ is the keplerian orbital frequency
\be
\label{kepler}
\omega_K=\dps{\sqrt{\frac{M}{R_0^3}}}~.
\ee
The source term can be written as
\be
\label{stressenergy}
T_{lm}(\o,r)=\delta(\o-m\o_K)\left[_0S^*_{lm}(\frac{\pi}{2},0)\,_0U_{lm}+
_{-1}S^*_{lm}(\frac{\pi}{2},0)\,_{-1}U_{lm}+\,_{-2}S^*_{lm}(\frac{\pi}{2},0)
\,_{-2}U_{lm}\right]\,.
\ee
The  functions $_{s}U_{lm}$ are given by
\beq
_0U_{lm}&=&\left(-2 \pi \sqrt{n(n+1)}~{\bar{\gamma}\Delta_0^2\over R_0^2}
\right)\delta(r-R_0),\nn\\
_{-1}U_{lm}&=&
\delta(r-R_0)\left(2\pi~\sqrt{2n}~m~\bar{\gamma}~\o_K^2~\Delta_0~ R_0^2\right)
-\delta'(r-R_0){\Delta^2\over r^2}
\left(2\pi\ii~\sqrt{2n}~\bar{\gamma}~\o_K ~R_0^2\right),\nn\\
_{-2}U_{lm}&=&
\delta(r-R_0)\left[\ii\pi~ m~\o_K^3~\bar{\gamma}~{\Delta_0^2\over R_0^2}~
\left(r^6\over\Delta\right)'_0~-\pi ~m^2~\o_K^4~\bar{\gamma}~R_0^6~\right]+\nn\\
&&+\delta'(r-R_0)\left[\Delta r^3\left(2\pi \ii~ ~m \o_K^3~\bar{\gamma}~ R_0\right)
+\Delta^2\left(4\pi~\o_K^2~\bar{\gamma}~R_0\right)\right]+\nn\\
&&+\delta''(r-R_0)\left[\Delta^2 r~\left(\pi~\o_K^2~\bar{\gamma}~R_0\right)\right],
\nn
\eeq
where the prime indicates differentiation with respect to $r$, and a subscript $0$
means evaluation at $r=R_0$.
As in Paper I, the solution of Eq. (\ref{teukolsky}) which satisfies  the boundary
conditions of  pure outgoing radiation at radial infinity, and  matches continuously
with the interior solution can be found by the Green's function technique, and
the amplitude of the wave emerging at radial  infinity can be shown to be
(cf. Paper I, Eqs. (4.3)-(4.7))
\be
\label{ampli0}
A_{l m}(\o)=
-{1\over W_{l m}(\o)}~
\int_{R}^{\infty}~ \frac{dr'}{\Delta^2}~\Psi^{~1}_{l m}(\o,r')~ T_{l m}(\o,r'),
\ee
where $W_{l m}(\o)$ is the Wronskian of the two independent solutions
of the homogeneous  BPT equation
\be
W_{l m}(\o)= \frac{1}{\Delta}\left[
\Psi^{~1}_{l m} \Psi^{~0}_{l m~,r}-\Psi^{~0}_{l m} \Psi^{~1}_{l m~,r}
\right],
\ee
and \op \Psi^{~0}_{l m}\cl and \op \Psi^{~1}_{l m}\cl 
satisfy the following equations
\be
\label{Ro}
\cases{\dps{
\IL_{BPT}\Psi^{~0}_{l m}(\o,r)=0}&\cr
\dps{
\Psi^{~0}_{l m}(\o,r\rightarrow\infty) =r^3e^{\ii\o r_*}}&\cr
},\qquad\qquad
\cases{\dps{
\IL_{BPT}\Psi^{~1}_{l m}(\o,r)=0} &\cr
\dps{\Psi^{~1}_{l m}(\o,R)= \bar{\Psi}_{l m}(\o,R)}&\cr
\dps{\Psi^{~1~'}_{l m}(\o,R)= \bar{\Psi}^{~'}_{l m}(\o,R)}&\cr
},
\ee
where $\IL_{BPT}$ is the differential operator on the left hand side
of the BPT equation, and \op \bar{\Psi}_{l m}(\o,R)\cl is the function 
constructed according to Eq. (\ref{relazione1}),  in terms of  the perturbed functions 
computed in the interior, and evaluated at the surface of the star.
The integral in Eq. (\ref{ampli0}) 
can explicitly be evaluated by making a
repeated use of the following property of the $\delta$-function
\[
\int g(r)\partial_r\left[ f(r)\delta(r-R_0)\right]~dr =
-g'(R_0)f(R_0).
\]
The result  is
\beq
\label{ampli1}
A_{l m}(\omega)&=&
- m_0~\frac{\delta(\o-m\o_K)}{W_{l m}(\omega)}
\left\{
\Psi^{~1}_{l m}(\omega,R_0)\left[
_0S^*_{l m}\left(-2\pi\sqrt{n(n+1)}~{\bar{\gamma}\over R_0^2}\right)\right.\right.\\
\nn
&+&
\left.\left.
_{-1}S^*_{l m}\left(
2\pi~\sqrt{2n}~m~\bar{\gamma}~\o_K^2~{R_0^2\over\Delta_0}
-4\pi\ii~\sqrt{2n}~\bar{\gamma}~\o_K~{1\over
R_0}\right)\right.\right.\\
\nn
&+&
\left.\left. 
_{-2}S^*_{l m}\left(2\ii\pi~
m~\o_K^3~\bar{\gamma}~{R_0^4\over\Delta_0^2}\left(
R_0-M\right)-\pi~ m^2~\o_K^4~\bar{\gamma}~{R_0^6\over\Delta_0^2}\right)
\right]\right.\\
\nn
&+&\left.
\Psi^{~1'}_{l m}(\omega,R_0)\left[
_{-1}S^*_{l m}\left(2\pi\ii~\sqrt{2n}~\bar{\gamma}~\o_K\right)+
_{-2}S^*_{l m}\left(-2\pi\ii m\o_K^3\bar{\gamma}{R_0^4\over\Delta_0}
-2\pi\o_K^2\bar{\gamma}R_0\right)
\right]\right.\\
\nn
&+&\left.
\Psi^{~1''}_{l m}(\omega,R_0)~
_{-2}S^*_{l m}
\left(\pi\o_K^2\bar{\gamma}R_0^2\right)
\right\}.
\eeq
If we define
\be
\label{ampli}
A_{l m}(\omega)
=\hat A_{l m}(\omega)\delta(\o-m\o_K),
\ee
the time-averaged energy-flux
\be
\label{enflux}
\dot E^R\equiv\left<dE_{GW}\over dt\right>=\lim_{T\rightarrow\infty}{E_{GW}\over T}
=\lim_{T\rightarrow\infty}{1\over T}\sum_{lm}\int d\omega
\left(dE_{GW}\over d\omega\right)_{lm}
\ee
can be written in terms of \op \hat A_{lm}(\o)\cl
as follows
\be
\label{amplispec}
\dot E^R (m\o_K)
=
\sum_{lm} \f{1}{4\pi (m \o_K)^2}~
\vert \hat A_{lm}(m\o_K)\vert^2
\equiv \sum_{lm}
\dot{E}^R_{lm}.
\ee
In order to evaluate $\Psi^{~0}_{l m}$ and 
$\Psi^{~1}_{l m}$, the BPT equation is integrated
with the same adaptive Runge-Kutta used for the equations in the interior.
We must remark that close to a resonance the solutions need to be
computed very accurately, since the Wronskian is the difference between two 
terms that almost cancel each other. When required, the tolerance in the integration
is decreased until convergence is reached.
Typically, computing the power emitted at a given
orbit takes about 10 s on a PC with a Pentium IV 1GHz processor.

Since the orbital frequency is related to the orbital velocity  $v$
and to the semilatus
rectum $p=R_0/M$ by the following relations
\be
v=\left( M\o_K\right)^{1/3}=\frac{1}{\sqrt{p}},
\ee
the energy flux \op \dot E^R \cl can also be considered as a function 
of $v$ or $p$.

In the following, we shall normalize \op \dot E^R \cl to the Newtonian quadrupole
energy flux
\be
\label{newtflux}
\dot E^N=\frac{32}{5}\frac{m_0^2}{M^2} v^{10}~.
\ee

\section{Comparing the gravitational flux of stars and black holes}

In our analysis, we consider five models of stars  with a polytropic EOS, 
using two values of the polytropic index,  \op n=1\cl and
\op n=1.5.\cl
The values of  the central density,  the
ratio \op\alpha_0\cl  between the energy density and the 
pressure at the centre, the mass,  the radius of the star, and the compactness
$M/R$ (in geometrical units) are given in Table \ref{table1}. 
We have chosen polytropic models for simplicity, but without loss
of generality, because 
the differences in the gravitational flux are expected to depend more
on global properties such as mass, radius, average density, or compactness,
rather than on the specific matter distribution. The parameters  we choose 
encompass a reasonable range of stellar models (radius ranging from 9 to 15 km), 
and the polytropic exponents, \op \Gamma \equiv 1+1/n=5/3\cl and \op \Gamma=2,\cl 
cover most of the range of structural properties obtained with realistic EOS's.

\subsection{Power radiated and resonant modes}

For each model, we integrate the equations of stellar perturbations 
as described in Sec. II, assuming that a point like mass,
$m_0,$ is moving on a circular orbit of radius $R_0$ with orbital
velocity \op v,\cl and we compute the energy flux emitted in gravitational 
waves normalized to the Newtonian quadrupole energy flux,
\op P \equiv \dot{E}^R/\dot{E}^N,\cl
where \op  \dot{E}^R\cl and \op \dot{E}^N\cl are given in Eqs.
(\ref{amplispec}) and (\ref{newtflux}).
In Ref.  \cite{poisson,poissonerratum,poisson2} the gravitational 
emission of a Schwarzschild black hole
perturbed by a  massive particle in circular orbit was studied numerically in
great detail.  Our results for the black hole  agree with them with an 
accuracy of (at least) one part in $10^6$. In order to allow an easy
comparison, and for future reference, we show in Table \ref{table1b} the 
contribution of the $2\le l \le 5$ multipoles to the total gravitational power.
The results for the black hole can directly be compared to Table II in Ref.
\cite{poisson2}.

\begin{figure}[htbp]
\begin{center}
\leavevmode
\epsfxsize=16cm \epsfbox{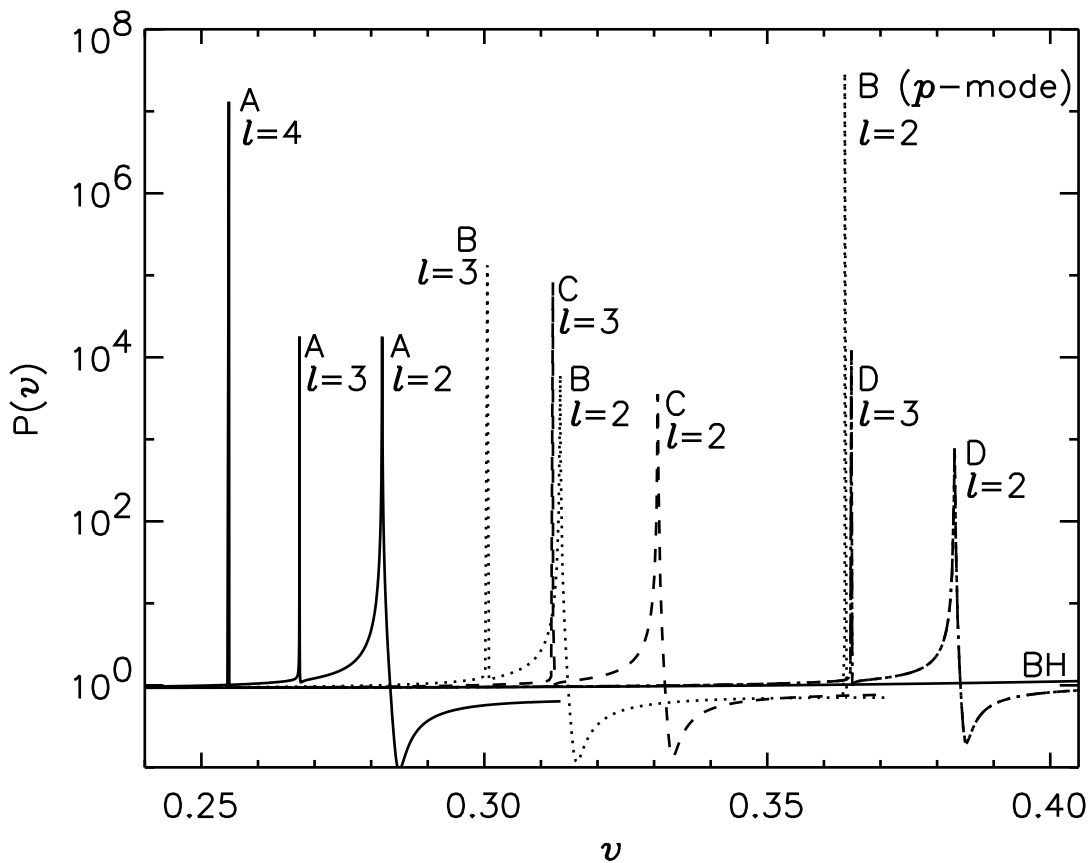}
\end{center}
\caption{
The normalized energy flux, $P(v),$ is plotted as a function of the orbital
velocity for the stellar models  given in Table \ref{table1} 
and for a black hole.
For model D and for the black hole the curves extend
up to the velocity which correspond to the ISCO, whereas for the other models 
they stop when the mass $m_0$ reaches the surface of the star.
The sharp peaks indicate that, for different values of the harmonic index $l,$
the fundamental quasi-normal modes of the star are excited if the orbital 
frequency satisfies the resonant condition (\ref{cond}); the curve of the 
stellar model B has a  peak at high $v$ which correspond to the
excitation of the first p-mode for $l=2.$
The most compact model E is not shown in the figure because at this scale
it is indistinguishable from the black hole.
}
\label{fig1}
\end{figure}

\begin{figure}[htbp]
\begin{center}
\leavevmode
\epsfxsize=16cm \epsfbox{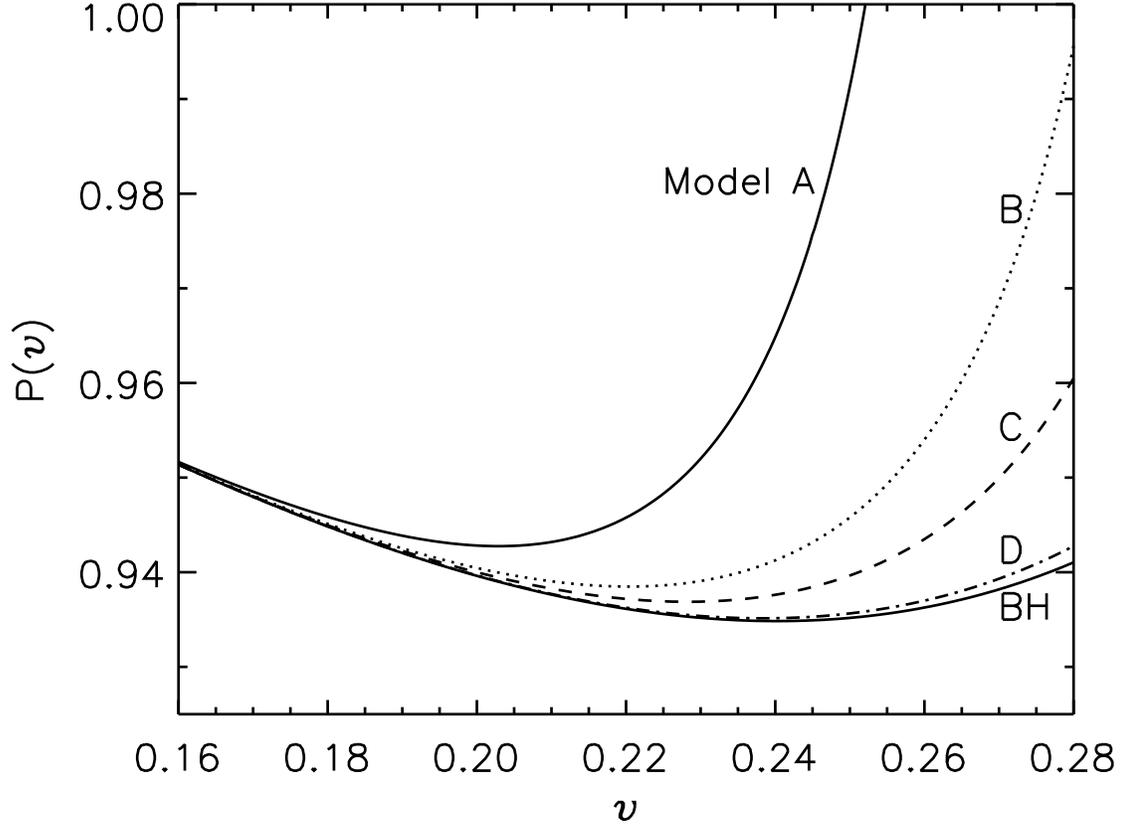}
\end{center}
\caption{
The normalized energy flux, $P(v),$ is plotted as in Fig. \ref{fig1}, but for
a smaller  orbital velocity  range, such that the peaks due to the excitation
of the stellar modes are excluded.
}
\label{fig2}
\end{figure}

In Fig.  \ref{fig1} we plot  the normalized 
energy flux, $P(v),$ as a function of the orbital
velocity, for the models of star we have considered, and for the black hole. 
$P(v)$ has been obtained by adding the contributions of different $l$'s and $m$'s, with
$2 \leq l \leq 7$. As discussed in paper I, a multipole of order $l$ contributes
to the total power as a correction of order $p^{2-l}$. By truncating our multipole
expansion at $l=7$ we are incurring in a relative error of order $p^{-6}=v^{12}$.
It should be mentioned that $P(v)$ is independent of the mass $m_0$.
For the black hole and  for the stellar model D, the plots
extend up to the velocity $ v=0.408,$ which corresponds to the Innermost Stable 
Circular Orbit (ISCO), \op R_0=6 M;\op for model A, B and C 
the ISCO would be  in the interior of the star, therefore
the plots are truncated at a smaller velocity, which corresponds to $m_0$ reaching 
the stellar surface.
Sharp peaks appear  if the central object is a star: they correspond to the
excitation of the fundamental quasi-normal modes of the star for different values of 
the harmonic index $l$.
In the case of model B the first p-mode for $l=2$ is also excited.
In Table \ref{table2} we show the values of the radius $R_0$, of the
dimensionless orbital velocity $v$, and of the  keplerian
frequency $\nu_K$ of the orbit that corresponds to the excitation of the 
fundamental modes of the star for different  $l$'s for the  considered stellar models.
The corresponding frequencies of the $f$-mode are given in the last column.
From the analytical form of the stress-energy tensor
(\ref{stressenergy}),  it is easy to see that, for each assigned $l$,
a mode of the star 
is excited when the orbital frequency
satisfies the resonant condition
\be
\label{cond}
m \nu_K =\nu_i
\ee
where $\nu_i$ is the mode frequency.  Table \ref{table2} shows that
the frequency of the f-mode increases with  $l$; however,  eq.
(\ref{cond}) shows that, for instance, the orbital velocity that corresponds to the 
excitation of the $f$-mode for 
$l=3$ is lower than that needed to excite the $f$-mode for
$l=2;$ this means that in the process of coalescence of the ``binary system" formed by 
the star and the point mass $m_0$, the  $f$-mode for $l=3$ is excited before the 
$l=2$ $f$-mode, and similarly the $l=4$ f-mode is excited before that for $l=3$,
while the gravitational wave frequency is higher.
The peaks corresponding to higher $l$ are narrower, so that 
peaks for $l>4$ are difficult to locate, even with the  use of an extremely 
refined grid.
From newtonian theory  we know that
the  $f$-mode frequency scales
with the square root of the average density of the star. 
This is true also in general relativity  and the
dependency on $\sqrt{M/R^3}$ is still linear \cite{asterosysm}.
This explains why for a chosen value of the polytropic index $n$ the peaks 
for  more compact stars occur at higher frequency, i.e. at higher $v$,
(for instance, compare in Fig.  \ref{fig1} the two curves for 
model B and D, for which  $n=1$  and $R=15$ km 
and $R=9.8$ km, respectively).
Since the  peaks corresponding to the mode excitation are very high,
the scale chosen on the vertical axis of Fig. \ref{fig1} makes
the response of the black hole to appear as a flat line.
The reason is that, since
the frequency of the lowest quasi-normal modes of a black hole are higher than those
of a star with the same mass,
the circular orbit that would excite them would have a radius smaller
than $6 M_{BH}.$ Thus, in the range of $v$ considered in Fig. \ref{fig1} the energy flux
emitted by the black hole is due essentially to the orbital motion. 
In Fig. \ref{fig2}, we show a zoom of Fig. \ref{fig1} 
restricted to the region $v < 0.28,$ which is  far enough from the
resonant orbits (except that for model A). In this case 
we can appreciate the differences between the emission of different stellar
models and that of a black hole.
If the orbital velocity is smaller than 0.16 all curves are practically
indistinguishable.

Fig. \ref{fig2} shows that the normalized energy fluxes emitted by
different stellar models have a different slope, and are always larger than the 
flux emitted by the black hole.
The curve for  model E ($n=1.5$, $R_0=9$ km) is practically
indistinguishable 
from the black hole  curve; that for  model D ($n=1$, $R_0=9.8$ km)
is also very close to  the black hole result.  At  first sight
we may relate this behavior to the fact that these stars are more 
compact than  models A,B,C; however,  the  steepest raise of the curves
of  models  A,B,C could also be a marginal effect of the resonances, or 
may be due to a different coupling between the orbital motion 
and the stellar structure.

\subsection{A harmonic oscillator model}

\begin{figure}[htbp]
\begin{center}
\leavevmode
\epsfxsize=12cm \epsfbox{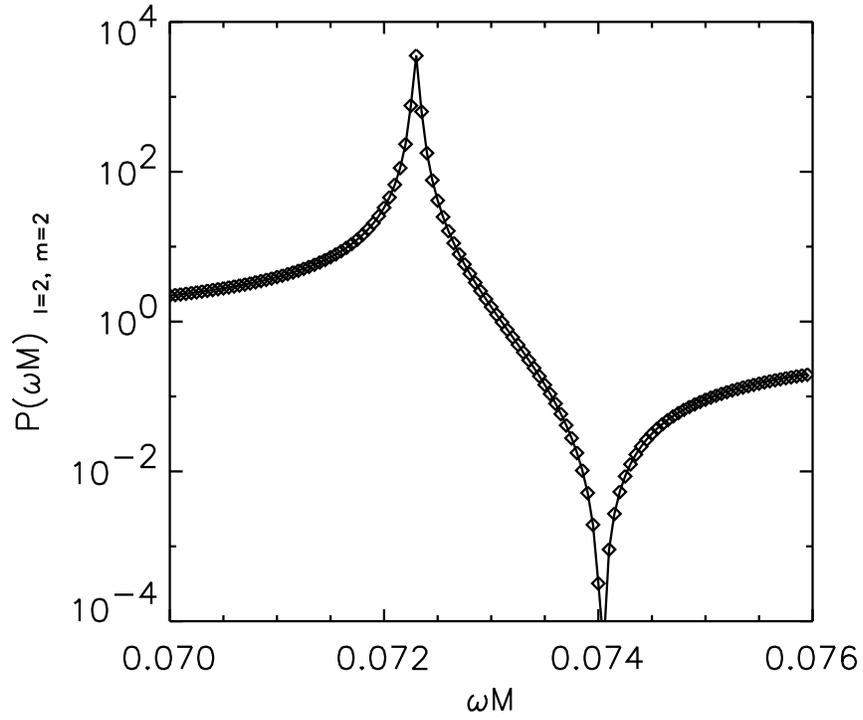}
\end{center}
\caption{
The normalized energy flux emitted by the stellar model C, is computed by using
the toy model (solid line) and by integrating the equations of
stellar perturbations for $l=2$ and $m=2$ (diamonds),
and plotted as a function of the dimensionless
orbital frequency $\omega M$ near the resonance of the $f-$mode.
The agreement is excellent.
}
\label{fig3}
\end{figure}

In order to better understand the underlying physical picture,
we shall use a toy model which has been employed in many contexts
to study the effects of stellar resonances (see for instance Ref. \cite{kojima}). 
Since a star oscillating in a quasi-normal mode of complex
frequency \op \omega_0-i\omega_i\cl
emits a gravitational wave of amplitude 
\op
\sim  C e^{\displaystyle -i(\omega_0-i\omega_i)t} \, ,
\cl
it can be modeled  as a harmonic oscillator
which satisfies the differential equation
\op
\ddot{X} + 2\omega_i \dot{X} + (\omega_0^2+\omega_i^2) X = 0 \, .
\cl
Here the amplitude $X$ is assumed to be normalized to some reference amplitude,
for instance to the amplitude of the newtonian quadrupole,
$A^N$, which we define in terms of 
the newtonian energy flux of Eq. (\ref{newtflux})  as follows
\be
\dot{E}^N=\frac{1}{4\pi \omega^2} \vert A^N (\omega)\vert^2.
\ee
The particle orbiting the star acts as a driving force  proportional to
$\omega^2 e^{-i\omega t}$, 
and the amplitude of the forced oscillation 
is found by  solving  the inhomogeneous equation
\be
\ddot{X} + 2\omega_i \dot{X} + (\omega_0^2+\omega_i^2) X=
b \omega^2 e^{-i\omega t}  \, ;
\ee
the solution can be written in the form
$ X(t) = X(\omega) e^{-i\omega t} \,$ , where
\be
\label{solu}
X(\omega) = \frac{-b \omega^2}{\omega^2 - \omega_0^2 -\omega_i^2 + 2 i \omega_i
\omega}.
\ee
We shall now  assume that, near a resonance, 
the total amplitude of the wave emitted by the perturbed star
is the sum of two contributions, one due to the orbital motion, the other to the
star pulsating in its quasi-normal mode, i.e.
\be
\label{nearres}
A_{tot}(\omega) =  A^N(\omega)\left[ 1+ X(\omega)\right].
\ee
According to this model, the normalized energy flux $P_{res}$  can be written as
\be
P_{res}(\omega)=\frac{|{A}_{tot}|^2}{\vert A^N\vert^2} = |1+X(\omega)|^2
=
\frac{[(1-b)\omega^2 - \omega_0^2 -\omega_i^2]^2 + [2\omega_i\omega]^2}
{[\omega^2 - \omega_0^2 -\omega_i^2]^2 + [2\omega_i\omega]^2}\, .
\label{resformula}
\ee
$ P_{res}(\omega)$ has a maximum in $\omega_0$ and a minimum in 
${\displaystyle \frac{\omega_0}{\sqrt{1-b}}} \,$ ;
the height of the maximum is 
\op {\displaystyle \frac{b^2 \omega_0^2}{4\omega_i^2}} \cl 
and that of the minimum is \op{\displaystyle
\frac{4\omega_i^2(1-b)}{b^2\omega_0^2}}\cl
(up to corrections of 
order  $O((\omega_i/\omega_0)^4)$).
For \op \omega << \omega_0,\cl $ P_{res}(\omega)$  tends to
$1$, while after the resonance it tends to $(1-b)$.
In order to check how good is the harmonic oscillator model to describe the behavior
of the star near a resonance, the quantity \op P_{res}(\omega)\cl
has to be compared  with the normalized energy flux, $P(\omega),$ computed by
integrating the equations of stellar perturbations.
In Fig. \ref{fig3} we plot
\op P_{res}(\omega)\cl (continuous line)
and the numerically evaluated  $P_{l=2, m=2}(\omega)$ (diamonds)  
as a function of  $\omega M,$
in a frequency region near the fundamental mode for $l=2, m=2,$ 
for the stellar model C.
The parameter $b$  is found from the locations of the maximum and the minimum, i.e.
$b=1-(\omega_{max}/\omega_{min})^2$, and $\omega_i$ is chosen in such a way that
the amplitude of the maxima of the two functions coincide.
The results are in very good agreement
(relative differences are of order $10^{-5}$)
except very close to the minimum, where the function $P_{l=2, m=2}(\omega)$ goes to zero. 
Incidentally,  it should be mentioned that
the evaluation of the exact value of the minimum for each $l$ is not so important, 
as far as the total energy flux is concerned; indeed, 
the total flux $P(\omega)$, evaluated at the frequency which corresponds
to a minimum for a given $l$, has a finite value, because it
is the sum of  contributions from  different
$l$'s and $m$'s, and it is dominated by the other multipoles.   

The simple analytical model
works surprisingly well in the whole range of each resonance,  capturing the
parabolic behavior in the region $|\omega - \omega_0|^2 \ll 1$,
used by several authors to calculate damping times, and
also describing the global behavior of the normalized flux with  high accuracy.
We found that the same procedure can satisfactorily be applied to
the other stellar models, and to resonances corresponding to
higher values of  $l$. 
The  values of the frequencies where the maximum  and the
minimum occur,  the  forcing amplitude $b$ and  the 
imaginary part of the frequency of the fundamental mode, \op \omega_i,\cl are
summarized in Table \ref{table3} for the five stellar models.
Note that the $\omega_iM$ scales approximately as $10^{-2l},$ and 
$b$ as $10^{-l}$.

\begin{figure}[htbp]
\begin{center}
\leavevmode
\epsfxsize=12cm \epsfbox{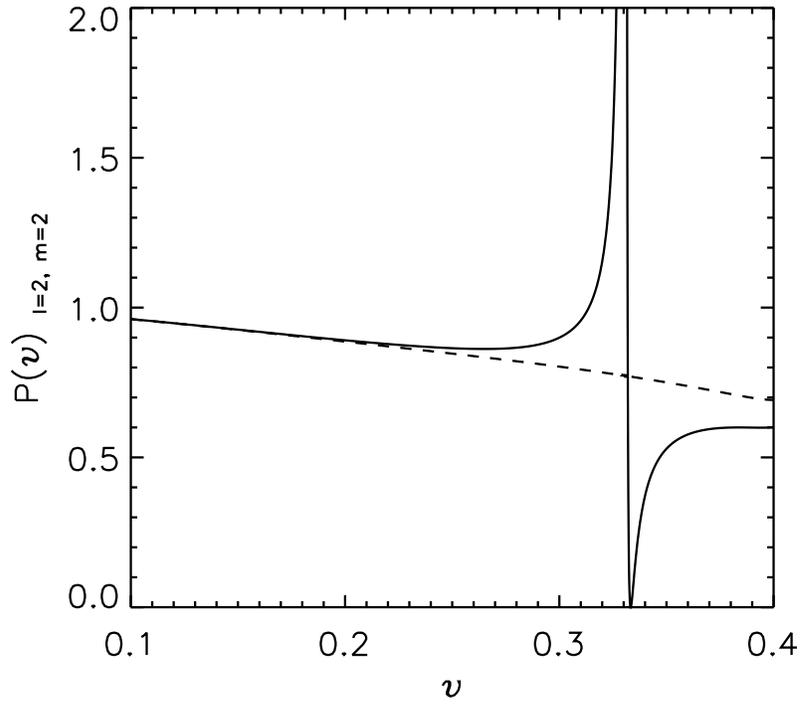}
\end{center}
\caption{
The $l=m=2$ contribution to the normalized energy flux, $P_{l=2, m=2}(v),$
is plotted as a function of the orbital velocity (solid line), and compared to
the result of {\it subtracting} the contribution of the resonance
(dashes) using the analytical model, as described in the text.
The data refer to the stellar model C.
}
\label{fig4}
\end{figure}

\begin{figure}[htbp]
\begin{center}
\leavevmode
\epsfxsize=16cm \epsfbox{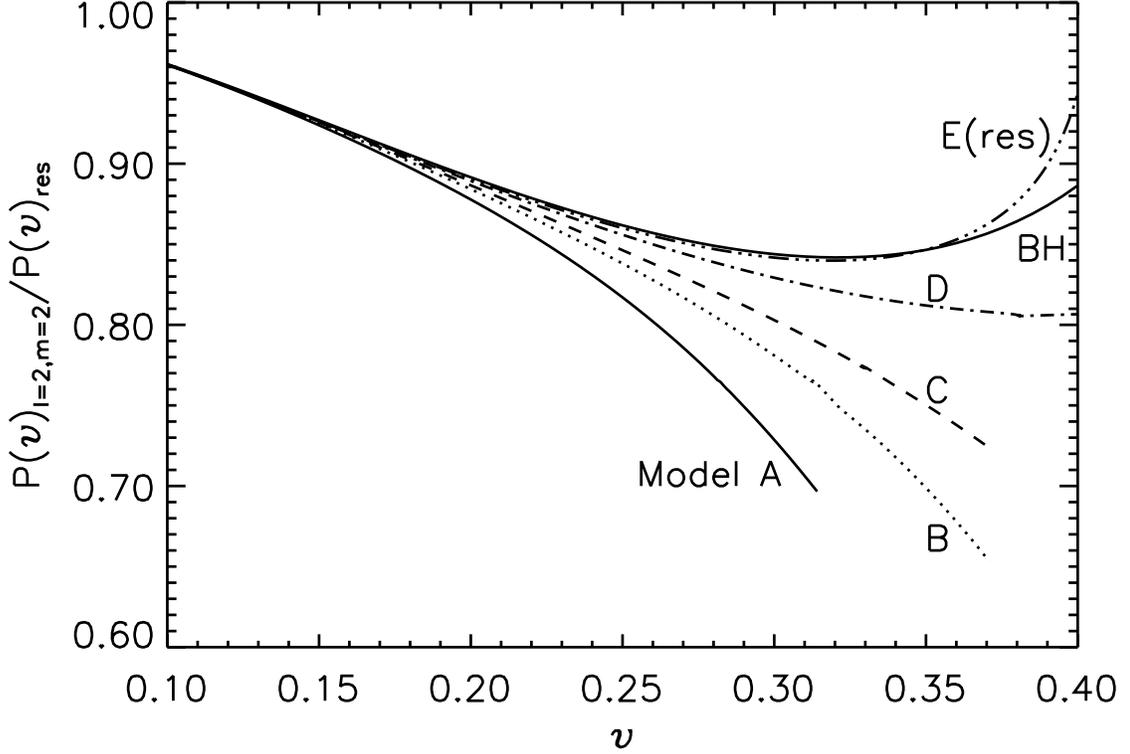}
\end{center}
\caption{In this figure we plot
the normalized  energy flux emitted by the four models of star for $l=2$ and $m=2$
versus the orbital velocity; the contribution of the resonances has been
removed by dividing $P_{l=2, m=2}(v)$ by $P_{res}$, as described in Sec. III.
For comparison, we plot the
normalized energy flux emitted by the black hole ($BH$)
and by the stellar model E ($E(res)$);
in this case we do not subtract the contribution of the resonant
excitation of the  quasi-normal modes because in order to excite these modes
the point particle should move on an orbit of radius smaller than 6M
(for the black hole), or smaller than the stellar radius (for model E).
}
\label{fig5}
\end{figure}

The agreement between the numerical result and the
toy model suggests the following
question: can the effects of a resonance be subtracted from
the global signal ?
The answer is shown in Fig. \ref{fig4}, where we plot the ($l=2,m=2$) contribution to the
flux, $P(v)_{l=2,m=2},$ with a solid line, and the
result of  subtracting the effect of the resonance with  dashed lines,
versus the orbital velocity.
The resonance-free curve has been obtained dividing the normalized energy flux 
emitted by the perturbed star by the model given in (\ref{resformula}), i.e.
$P(v)_{l=2,m=2}/P_{res}(v)$. 
Now the question of whether or not the different slope of the curves plotted 
in Fig. \ref{fig2} can be attributed to the effect of resonant excitation of a
stellar quasi-normal mode can be answered positively. The increase in the 
energy output at orbital velocities of about $v=0.2$ is just an effect of the
resonance tail. 
However, it is not the only effect.
Indeed, since we are normalizing the emitted flux to the newtonian quadrupole flux, 
if the signal emitted by the perturbed star would be, as we assume in 
eq. (\ref{nearres}), the sum of a term due to the orbital motion (the quadrupole) and 
a term due to the resonance, the dashed line in 
Fig.  \ref{fig4}  should be a horizontal straight line.  Conversely,
it is a slightly decreasing function of  $v$.
In order to see if this is a general feature, we  have also calculated 
$P(v)_{l=2,m=2}/P_{res}(v)$
for the stellar models A, B and D. The results are plotted  in Fig. \ref{fig5}.
For comparison, we also plot $P(v)_{l=2,m=2}$ for a black hole and for 
the stellar model E, computed  without subtracting the contribution of the resonances.
The reason is that we cannot subtract this contribution  because,
 as mentioned before, black hole quasi-normal modes cannot be
excited before the ISCO and similarly, to excite the f-mode of 
model E  the point mass should move on an orbit with radius smaller than the
stellar radius.
We see that after the subtraction there is still a
difference between different EOS's; in particular, if we classify
the models A,B,C,D according to their stiffness (where a stiffness indicator is,
for example, the speed of sound at a given density), we find that 
the stiffer the EOS is, the smaller is the orbital contribution to the total
emission. We must remark, however, that when  the effect of the
resonances is included, this trend is inverted; 
stiffer models emit more energy, because their resonant frequencies are smaller 
and their effect is more pronounced.

The influence that the structural effects discussed above may have on the detection
of these signals  by terrestrial detectors,
should be estimated by computing the phase evolution of the signal; this can be done 
by taking into account the evolution of the orbit
due to radiation reaction effects. Although this problem will be 
specifically addressed in a separate paper, we can anticipate that,
depending on the stellar model,  most of the phase shift is accumulated during 
the last 50 cycles before merging; this effect is not likely to be detectable 
by the first generation of LIGO and VIRGO detectors, unless we are so lucky that a 
NS-NS coalescence occurs closer than 10 Mpc.
However, the situation would be different if a detector like EURO
\cite{euro}, recently proposed, would be constructed, because it would be 
very sensitive in the frequency range (600-1200) Hz
 where the structural effects  become significant.
Quantifying precisely the differences between different signals and templates
is out of the scope of this paper, and requires a careful calculation of 
the overlaps of the "true signal" with those that are currently used as templates. 
This work is in progress and will be reported in a forthcoming paper.

\section{On the validity of the perturbative approach}
\begin{figure}[htbp]
\begin{center}
\leavevmode
\epsfxsize=16cm \epsfbox{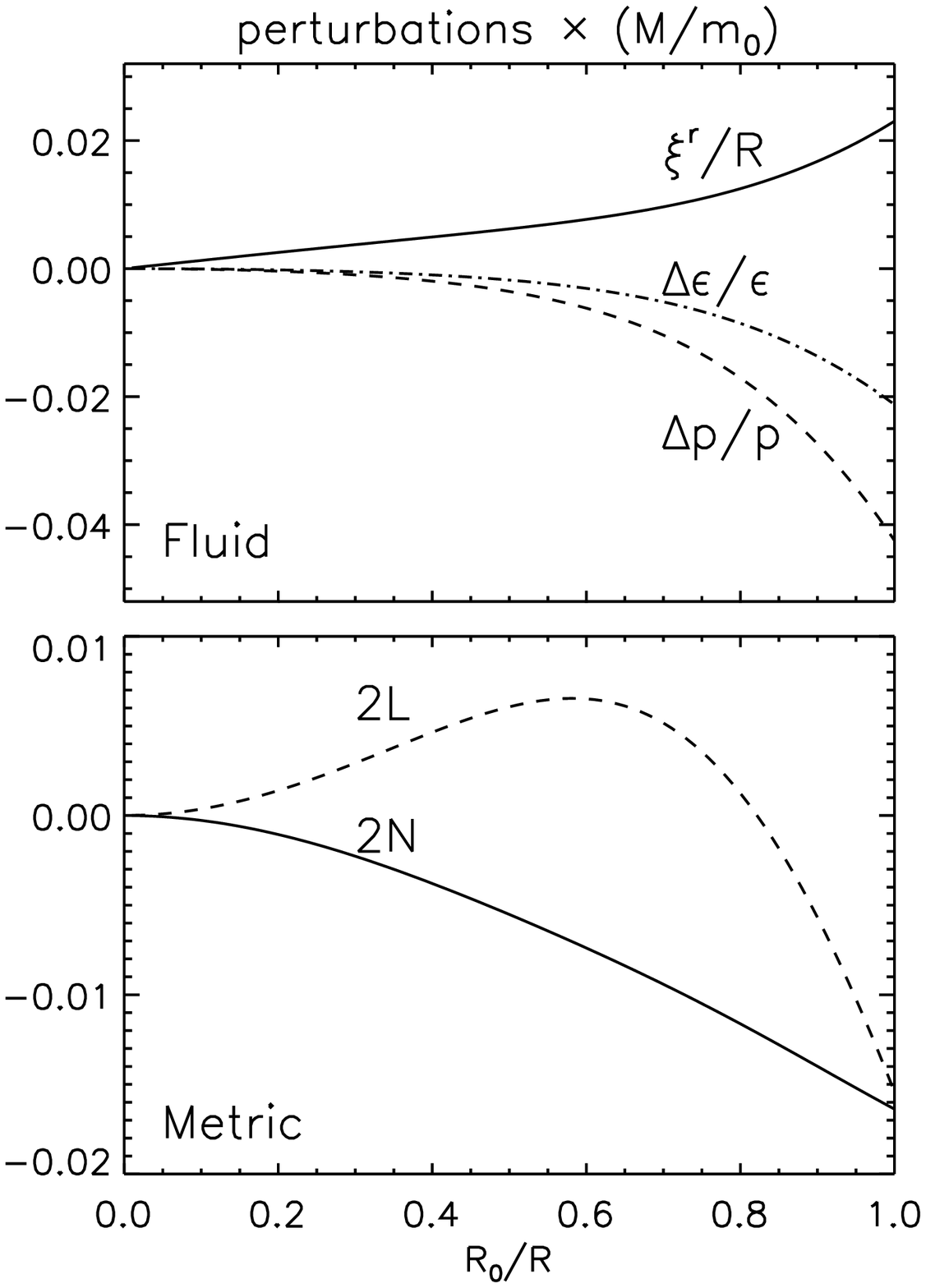}
\end{center}
\caption{
This figure refers to the stellar model D perturbed by a point mass moving on a
circular orbit of radius $R_0=3R$.
All quantities are normalized to the ratio $m_0/M$, and only the $l=2,m=2$
contribution is shown, which is the most significant.
In the upper panel, the radial component of the lagrangian displacement
normalized to the radius of the star, $\xi_r(r)/R,$  and the
lagrangian perturbations of the density and of the pressure 
normalized to their equilibrium values,
$\Delta p(r)/p(r)$ and $\Delta \epsilon(r)/\epsilon(r),$ are plotted as a function
of the radial distance from the center of the star.
The metric perturbations $2N$ and $2L$ plotted in the lower panel,
have to be compared to  unity, since they are a measure of
the deviations of the perturbed metric functions $\nu(r)$ and
$\mu_2(r)$ with respect to their equilibrium values (cf. Paper I, Eq. (2.1)).
}\label{fig6}
\end{figure}

The theory of perturbations of black holes  excited by a point particle developed
since the early seventies was based on the assumption that the mass  
of the particle is much smaller than the  black hole mass;
under this condition, the effect of the particle on the black hole
is that of inducing  a small perturbation on the equilibrium configuration,
and the stress energy tensor of the mass $m_0$ can be considered as
a source for the perturbed equations. 
In addition, since $m_0$ does not affect the
background geometry, it will move on a geodesic of the unperturbed spacetime.
The same assumption can be done to study the 
perturbations of a star, as we do in this paper. 
However, since one of the purposes of our study is to have an insight on the
last phases of the coalescence of a true binary system composed of stars of
comparable mass, we may ask the following question: how big can  the mass 
$m_0$ be, in order the deviation it induces on the gravitational field and on
the thermodynamical structure of the companion star to be considered as a
perturbation? Or, we can formulate the same  question in a different way: 
given a mass $m_0$, not necessarily small, moving  on a circular
orbit around a star, up to which distance its effect on the
companion can be considered as a perturbation?
\footnote{
We neglect the fact that if $m_0$ is large it will not move on a geodesic of the
unperturbed spacetime. This point will be discussed in the concluding remarks.}
To answer this question in Fig. \ref{fig6} we plot the  fluid and the gravitational 
perturbed functions inside the star, for $l=2, m=2$ (the most significant
contribution) and for the stellar model D, 
assuming that the point mass is 
moving on a circular orbit at a distance $R_0=3 R$ from the star.
In the upper panel we plot the radial component of the
Lagrangian displacement normalized to the 
stellar radius and the Lagrangian
perturbations of the density and of the pressure normalized to their equilibrium
values.
In the lower panel we plot the perturbations of the metric
$2N(r)$ and $2L(r)$, that have to be compared  to unity (cf. Paper I, Eq. (2.1)). 
All quantities are normalized to the ratio $m_0/M$. 
Similar results are  obtained for the other stellar models.
Fig. \ref{fig6} shows
that, even if we assume that the two masses are equal, all perturbations
belonging to the fluid or to the gravitational field
are small compared to the corresponding unperturbed functions. 
Thus, the perturbative approach  holds  even when the two bodies
have comparable mass and are as close as three stellar radii. 
Corrections of the same order of magnitude have been found in full nonlinear
stationary solutions of binary neutron stars \cite{gourg}

\section{Concluding remarks}
In this paper we have studied in a full general relativistic approach
how the  internal structure of a neutron star affects its gravitational 
emission when the star is perturbed by a  close, orbiting companion.
Our study has been done by using a perturbative approach in 
the frequency domain, and 
the massive object which perturbs the star has been assumed to be a point particle
of mass $m_0$. We have considered five models of star with global properties  - like
compactness, average density, stiffness of the EOS - that encompass a wide range 
of stellar properties, and we have compared their behavior with that of a black hole
excited by the same  source.

The results we find can be summarized as follows.
The internal structure of the star affects the emitted flux of gravitational waves 
only when the orbital velocity of $m_0$ is (approximately)
higher than one fifth of the speed of light, i.e. when  $m_0$ 
is very close to the central star.
If the mass of the star and  $m_0$ were  comparable and about $1.4~M_\odot,$
this velocity would correspond to an emission frequency  $\gappreq 185$ Hz.
For lower values of $v$ (larger orbital separation)
neutron stars and black holes behave in the same manner and
the gravitational fluxes they emit are  practically  indistinguishable. 

The difference in the energy flux emitted by a star and a black hole arises
mainly  because of the excitation of the fundamental mode of the star for 
different $l$'s, which can be
modeled extremely well in terms of a suitably defined harmonic oscillator.
The results for model A suggest that the emission properties of low mass neutron
star binary systems would deviate significantly from those of black hole-black hole
binary systems.

Once the effect of the resonant excitation of the quasi-normal modes of the star are
subtracted, we find that there is still a residual difference between the emission 
of neutron stars and black holes, and that stars with stiffer EOS emit less energy
than a black hole. However, this trend is inverted when the effect of resonances 
is included.

It should be stressed that the region where corrections due to the EOS 
begin to be significant is the same region where high order Post Newtonian (PN) 
corrections play a significant role. Thus these effects should be included in constructing 
templates  to be used in the data analysis of gravitational detectors.
In addition, the corrections introduced by the tail of the resonant f-mode
is  comparable, if not larger, to that introduced by the  high order PN corrections 
\cite{damouriyersathia}.

The problem of mode excitation during the latest phases of the evolution of a binary
system, and the consequences it has on the emitted gravitational signal,
have previously been studied by Ho \& Lai \cite{HoLai} by a newtonian approach.
In particular, as far as the excitation of the f-mode is concerned, they consider 
the case of a rotating,  incompressible model of star, which is very close to our model E, 
whereas we use  a relativistic perturbative approach and focus on the 
differences which may arise because of different EOS's.
Apart from this, we basically reach similar conclusions about the 
order of magnitude of the resonant effects and on the regime where they may play a role.

It is interesting to discuss to what extent
can we extrapolate our results to the case when the mass $m_0$ is comparable to
that of the central star, i.e, to simulate a true coalescing binary system. 
Indeed, the results of Sec. IV indicate that the perturbative approach 
holds also when  two bodies of comparable mass are very close, 
even closer than 3 stellar radii. 
A naive generalization of  the energy fluxes we obtain
could be that of  rescaling the amplitude of the energy flux by the appropriate
value of $m_0$, and  to change
the orbital frequency $\omega_K=\sqrt{M/R_0^3}$ 
(and consequently the emission frequency $\omega_{GW}=m\omega_K$) 
by replacing the mass of the star $M$ with the total mass
of the system $M_t=M+m_0;$  it is worth mentioning that,
in this case, given a certain orbital radius $R_0$, the orbital
frequency corresponding to that radius would be higher than that of the point
particle on the same orbit. 
This can easily be done, but the results should be considered
only as an indication of what may happen in reality. Indeed,
in order to correctly generalize the results,  the perturbative approach
should be improved in many ways: 
the geodesic equation which are used to describe the motion of  $m_0$ 
should be replaced by  the correct equations of motion of a two-body system, 
as seen in the coordinate system centered on 
one star; these equations are now known  at the 3.5 PN
level \cite{blanchet}, and they would change the stress energy 
tensor we put on the right-hand side of our equations;  
the problem of chosing the reference frame in which to compute the radiated power as
seen by a distant observer is a very delicate one, and in order to compare the results 
of the perturbative approach for equal masses with those of the post-newtonian approach 
one should change to a frame centered in the center of mass of the binary system;
in computing the orbital evolution due to radiation reaction effects it is customary
to use the adiabatic approximation which assumes that the timescale of the orbital
evolution is larger than the orbital period; we think that the role of this assumption
has to be investigated very carefully during the last few cycles before coalescence.
All of these problems are important:  we are looking for very small effects 
which  arise just before coalescence, and  they may introduce corrections of a
few percents in the emitted power, which are of the same order  as
the effects of stellar structure we discuss in this paper. 
Finally, one should certainly consider rotating stars.
Rotation may change the situation in  a positive
direction, because we know that it has the effect of lowering some
of the mode frequencies, and this may enhance the marginal effect of the mode excitation
which we see in the tail of the signals emitted by compact stars.

In future papers we plan to implement all these effects in our perturbative approach, with
the final goal of providing accurate templates which may be used in the data analysis of future
high sensitive detectors operating in the kHz region.

\acknowledgments
We are indebted to the anonymous referee for a number of useful comments
and suggestions.
This work has been supported by the EU Programme 'Improving the Human
Research Potential and the Socio-Economic Knowledge Base' (Research
Training Network Contract HPRN-CT-2000-00137).

\begin{table}
\centering
\caption{
Parameters of the polytropic stars we consider in our analysis:
the polytropic index $n$, the central density, the ratio 
$\alpha_0=\epsilon_0/p_0$ of central energy
density to central pressure, the mass,  the radius  and the ratio $M/R$
($\alpha_0$ and $M/R$ are in geometric units).  The central energy density 
is chosen in such a way that the stellar mass is equal to
$1.4 M_\odot$, except for model A, the mass of which is about one solar mass. 
}
\vskip 12pt
\begin{tabular}{@{}clllllr@{}}
\hline
Model number &$n$ &$\rho_c$ (g/cm$^3$) &$\alpha_0$  &$M$ ($M_\odot$) &$R$ (km)&
$M/R$\\
\hline
A   &1.5     & $1.00\times 10^{15}$    &13.552     &$0.945$     &14.07 & 0.099\\
B   &1       & $6.584\times 10^{14}$   &9.669      &$1.4$       &15.00 & 0.138\\
C   &1.5     & $1.260\times 10^{15}$   &8.205      &$1.4$       &15.00 & 0.138\\
D   &1       & $2.455\times 10^{15}$   &4.490      &$1.4$       &9.80  & 0.211\\
E   &1.5     & $8.156\times 10^{15}$   &2.146      &$1.4$       &9.00  & 0.230\\
\hline
\end{tabular}
\label{table1}
\end{table}

\begin{table}
\centering
\caption{
Decomposition of the gravitational luminosity $(M/m_0)^2 \dot{E}^R$
into multipole contributions for $p=10$. The BH results can be compared directly
to those of Table II in Ref. [10].
}
\vskip 12pt
\begin{tabular}{@{}cccccccc@{}}
\hline
$l$ & $m$ & BH & A & B & C  & D & E \\
\hline 
2   & 2 & 0.536879(-4) & 0.339696(-04) & 0.220321(-06) & 0.667680(-04) & 0.542028(-04) & 0.537526(-04)\\
2   & 1 & 0.193161(-6) & 0.194489(-06) & 0.193693(-06) & 0.193491(-06) & 0.193244(-06)& 0.193189(-06)\\
3   & 3 & 0.642607(-5) & 0.588917(-05) & 0.599186(-05) & 0.564656(-05) & 0.643160(-05)& 0.642658(-05)\\
3   & 2 & 0.479591(-7) & 0.481018(-07)& 0.479889(-07) & 0.479742(-07) & 0.479605(-07)& 0.479594(-07)\\
3   & 1 & 0.571489(-9) & 0.712057(-09) & 0.587559(-09) & 0.577923(-09) & 0.571763(-09)& 0.571520(-09)\\
4   & 4 & 0.953958(-6) & 0.918437(-06)& 0.944728(-06) & 0.948295(-06) & 0.954066(-06)& 0.953966(-06)\\
4   & 3 & 0.877874(-8) & 0.879191(-08)& 0.878020(-08) & 0.877939(-08) & 0.877878(-08) &  0.877876(-08)\\
4   & 2 & 0.526223(-9) & 0.116026(-08)& 0.532315(-09)  & 0.528287(-09) & 0.526252(-09)& 0.526227(-09)\\
4   & 1 & 0.145758(-12) & 0.145966(-12) & 0.145781(-12) & 0.145768(-12) & 0.145759(-12) & 0.145759(-12)\\
5   & 5 & 0.152415(-6) & 0.149988(-06)& 0.152100(-06)  & 0.152269(-06) & 0.152418(-06)& 0.152415(-06)\\
5   & 4 & 0.149211(-8) & 0.149338(-08) & 0.149219(-08) & 0.149214(-08) & 0.149212(-08)& 0.149212(-08)\\
5   & 3 & 0.182910(-9) &  0.168856(-09) & 0.184547(-09) & 0.183313(-09) & 0.182911(-09)& 0.182910(-09)\\
5   & 2 & 0.381934(-12) & 0.382239(-12) & 0.381952(-12) & 0.381942(-12) & 0.381935(-12) & 0.381935(-12)\\
5   & 1 & 0.236763(-15) & 0.252126(-15) & 0.237281(-15) & 0.236943(-15) & 0.236765(-15)& 0.236764(-15)\\
\hline
\end{tabular}
\label{table1b}
\end{table}

\begin{table}
\centering
\caption{
In this table we give the values of the radius, orbital velocity and keplerian 
frequency $(\nu_K)$
of the circular orbits which correspond
to the excitation of the fundamental mode of the considered stars
for the first relevant multipoles, whose frequency is given in the last column.
For model E we do not give these data for $l=2$ because  in order to excite the 
corresponding mode $R_0$ should be smaller than the ISCO $(6M)$.
}
\vskip 12pt
\begin{tabular}{@{}llllll@{}}
\hline
Model   &$l$  &$R_0$ (km) &$v$    &$\nu_K$ (Hz)       &$\nu_f$ (Hz)       \\
\hline
A       &4     &31.8     &0.255  &567            &2260            \\
        &3     &29.0     &0.267  &651            &1953         \\
        &2     &26.0     &0.282  &767            &1534         \\
\hline
B       &3     &22.9     &0.300  &626            &1879             \\
        &2     &21.0     &0.313  &711            &1422            \\
\hline
C       &3     &21.2     &0.312  &702            &2105             \\
        &2     &18.9     &0.331  &835            &1671             \\
\hline
D       &3     &15.5     &0.365  &1119           &3358            \\
        &2     &14.1     &0.383  &1296           &2593           \\
\hline
E       &3     &13.5     &0.391  &1379           &4138            \\
\hline
\end{tabular}
\label{table2}
\end{table}

\begin{table}
\centering
\caption{ 
Parameters of the resonances (\ref{resformula}) corresponding to
the fundamental mode of the most relevant $l$-multipoles. All frequencies
are normalized to the mass of the corresponding  star.
}
\vskip 12pt
\begin{tabular}{@{}llllll@{}}
\hline
Model   &$l$  &$(\omega M)_{max}$  & $(\omega M)_{min}$  &$b$  & $\omega_i M$       \\
\hline
A       &4     &0.066218  &0.066239  &0.000645  &9.67$\times
10^{-10}$ \\
        &3     &0.057314  &0.057495  &0.006272  &1.32$\times 10^{-7}$ \\
        &2     &0.044823  &0.046273  &0.061682  &1.81$\times 10^{-5}$ \\
\hline
B       &3     &0.081420  &0.081645  &0.005618  &1.93$\times 10^{-7}$ \\
        &2     &0.061581  &0.063363  &0.055367  &1.90$\times 10^{-5}$  \\
\hline
C       &3     &0.091210  &0.091402  &0.004208  &2.09$\times 10^{-7}$ \\
       &2     & 0.072299  &0.074036  &0.046373  &2.47$\times 10^{-5}$ \\
\hline
D       &3     &0.145709  &0.145895  &0.002555  &6.17$\times 10^{-7}$ \\
        &2     &0.112454  &0.114287  &0.031836  &5.77$\times 10^{-5}$   \\
\hline
E       &3     &0.179606  &0.1796935 &0.000971  &  4.91$\times 10^{-7}$  \\
\hline
\end{tabular}
\label{table3}
\end{table}



\end{document}